\documentclass[english,aps,floats,onecolumn,showpacs,nofootinbib]{revtex4}
\usepackage{pslatex}
\usepackage[T1]{fontenc}
\usepackage[latin1]{inputenc}
\usepackage{graphicx}
\usepackage{epsfig}

\usepackage{calc}
\usepackage{ifthen}

{
{
{
\newcommand{\bea}{\begin{eqnarray}}
\newcommand{\eea}{\end{eqnarray}}

\newcommand{\nc}{\newcommand}
\nc{\renc}{\renewcommand}
\nc{\eqs}[2]{\mbox{Eqs.~(\ref{#1},\,\ref{#2})}}
\nc{\eq}[1]{\mbox{Eq.~(\ref{#1})}}
\nc{\figs}[2]{\mbox{Figs.~(\ref{#1},\,\ref{#2})}}
\nc{\fig}[1]{\mbox{Fig~.(\ref{#1})}}
\nc{\be}[1]{\begin{equation} \mbox{$\label{#1}$}}
\nc{\ee}{\vspace{0.1cm}\end{equation}}

\newcommand{\bean}{\begin{eqnarray*}}
\newcommand{\eean}{\end{eqnarray*}}

\def\GeV{{\rm \ GeV}}
\def\MeV{{\rm \ MeV}}

\def\TeV{{\rm \ TeV}}

\def\bfk{{\bf k}}

\def\tM{\tilde{M}}

\def\vphi{\varphi}

\def\lae{\;^{<}_{\sim} \;} \def\gae{\; ^{>}_{\sim} \;}

\def\stau{\tilde{\tau}}

\begin{document}
\title{Dark Matter Gravitinos and Baryons via Q-ball decay in the Gauge-Mediated MSSM}
\author{Francesca Doddato}
\email{f.doddato@lancaster.ac.uk}
\author{John McDonald}
\email{j.mcdonald@lancaster.ac.uk}
\affiliation{Lancaster-Manchester-Sheffield Consortium for Fundamental Physics, Cosmology and Astroparticle Physics Group, Dept. of Physics, University of 
Lancaster, Lancaster LA1 4YB, UK}
\begin{abstract}

    We show that late Q-ball decay in the MSSM with gauge-mediated SUSY breaking can provide a natural source of non-thermal NLSPs which subsequently decay to gravitino dark matter without violating nucleosynthesis constraints. To show this, we perform a global analysis of Q-ball formation and decay in Affleck-Dine baryogenesis for a $d = 6$ $(u^{c}d^{c}d^{c})^2$ flat direction of the gauge-mediated MSSM.  A general phenomenological potential for the flat-direction is studied and the Q-ball decay properties are obtained as a function of its parameters. The corresponding gravitino mass necessary to account for dark matter is then determined for the case of stau NLSPs. The decay temperature depends on the charge of the Q-balls, which is determined by the fragmentation of the AD condensate. Different fragmentation scenarios are considered, and the final non-thermal NLSP density from Q-ball decay and NLSP annihilation is determined. Particular care is taken to establish that NLSPs from Q-ball decay become homogeneous and non-relativistic prior to annihilation. The gravitino mass necessary for dark matter is naturally consistent with the theoretical gravitino mass in the gauge-mediation model.

\end{abstract}
\pacs{  }
\maketitle

\section{Introduction}

         A fundamental question in cosmology is the nature of dark matter (DM) and, in particular, the identity of the dark matter particle in the case where it has a particle physics origin. In the context of the MSSM and its extensions, the dark matter candidate is fundamentally connected to the nature of SUSY breaking. In gravity-mediated SUSY breaking the neutralino is a natural lightest SUSY particle (LSP) candidate. In gauge-mediated SUSY breaking (GMSB), the gravitino becomes a natural LSP candidate.

     There are two ways a gravitino DM density might be produced in GMSB models. For sufficiently large reheating temperature ($T_{R} \sim 10^{7} \GeV$ for the case $m_{3/2} \sim 0.1 \GeV$), gravitinos can be produced via thermal scattering \cite{gravth}. Alternatively, gravitinos could be produced via the late decay of next-to-lightest supersymmetric particles (NLSPs). The latter process is, however, strongly constrained by big-bang nucleosynthesis (BBN) \cite{kazbbn}. Decay of conventional thermal relic NLSPs to a DM gravitino density is already completely excluded by BBN. Therefore if gravitino dark matter comes from NLSP decay, a source of {\it non-thermal} NLSPs is necessary. Even in this case, the mass of the gravitino must be below 
1 GeV in order to satisfy BBN constraints on NLSP decay. The weakest BBN upper bounds are obtained for the case of stau and sneutrino NLSPs, which can allow gravitino masses up to 0.1-1 GeV \cite{kazbbn}. 

               A second fundamental question is the origin of the baryon asymmetry. In the MSSM a simple and natural mechanism for generating the baryon asymmetry is Affleck-Dine baryogenesis \cite{ad}. Remarkably, this can also provide a source of non-thermal NLSPs. 
This is because the AD condensate can fragment to Q-balls which subsequently decay to NLSPs
at a low temperature \cite{fd1,fd2}. For the case of $d=6$ flat directions, the Q-balls will typically decay to a density of NLSPs at a temperature below 1 GeV, which is well below the NLSP freeze-out temperature\footnote{Alternative Q-ball decay scenarios, in which the Q-balls are assumed to decay directly to gravitino dark matter rather than to NLSPs, are presented in \cite{shoeQ,kkQ}.}. 

   Thus, without adding any fields beyond those in the MSSM,
it is possible to simultaneously produce both the baryon asymmetry and gravitino dark matter via AD baryogenesis.  In fact, only Q-ball decay, through the production of a non-thermal NLSP density, can produce gravitino dark matter in the case of AD baryogenesis along a d = 6 flat direction\footnote{It may also be possible to produce gravitino dark matter via inflaton decay \cite{infgrav} or decay of the fields associated with SUSY breaking \cite{modgrav}, but this is highly dependent upon the model of inflation and SUSY-breaking.}.  This is because the reheating temperature in $d = 6$ AD baryogenesis must be below about 100 GeV in order to produce the observed baryon asymmetry \cite{fd1}, which is too low for thermal generation of gravitino dark matter.

    Before summarizing the analysis to be carried out here, 
we first state our main result. We will show that it is possible to produce gravitino dark matter via decay of the non-thermal NLSPs coming from Q-ball decay. The required dark matter gravitino mass can be light enough to be consistent with BBN constraints on NLSP decay to gravitinos. Moreover, the gravitino mass required to explain dark matter is naturally consistent with the mass predicted by the GMSB model. Therefore $d = 6$ AD baryogenesis in GMSB can produce both gravitino dark matter and the baryon asymmetry without violating BBN constraints. This model can therefore simultaneously solve two fundamental problems in the context of MSSM cosmology with GMSB: the origin of the baryon asymmetry and the origin of dark matter.

     For the case of a $d = 6$ flat direction, we showed in \cite{fd1} that if the messenger mass scale is sufficiently close to the magnitude of the AD field $\Phi$ at the onset of baryogenesis, AD condensate fragmentation and the resulting Q-balls will typically occur near the region of the flat-direction potential between the logarithmic GMSB "plateau" at large $|\Phi|$ relative to the messenger mass and the quadratic potential at small $|\Phi|$. In this case the Q-balls will be unstable and can decay before BBN. For this to occur, the messenger mass scale must be large enough, which in turn requires that $m_{3/2}$ is not much smaller than 1 GeV. In this case the Q-balls can have properties which are intermediate between the conventional gauge-mediated and gravity-mediated Q-balls\footnote{By conventional gauge-mediated Q-balls, we mean Q-balls in which $|\Phi|$ is large enough that the solutions are completely dominated by the logarithmic potential.}. 
In \cite{fd2} we studied the properties of GMSB Q-ball solutions in detail, in particular how they interpolate between gravity-mediated and gauge-mediated type Q-balls and how their decay temperature depends on the potential parameters.

   Our previous studies established the nature of the initial fragmentation of the AD condensate and the form of Q-ball solutions in GMSB models.  In this paper we consider whether the combined effect of fragmentation to Q-balls and the subsequent decay of these Q-balls to NLSPs can explain gravitino DM while remaining consistent with BBN constraints. We will combine the semi-analytical fragmentation model of \cite{fd1} with the numerical Q-ball solutions of \cite{fd2}. Using these in a global analysis of AD baryogenesis, condensate fragmentation and Q-ball formation, we will be able to establish the Q-ball decay temperature to NLSPs and so determine the gravitino density and dark matter gravitino mass from the subsequent annihilation and decay of the NLSPs.

    Our analysis is based on a phenomenological potential, first introduced in \cite{fd1}, which models a general flat-direction potential in the MSSM with gauge-mediation. This potential is characterized by a number of parameters, in particular the soft scalar mass $m_{s}$, 
the messenger mass $M_{m}$ (above which SUSY breaking is suppressed), the GMSB A-term $a_{o}$ and the mass scale of the non-renormalizable operators responsible for AD baryogenesis, $\tM$. Using this potential, we will evolve the flat-direction condensate through
AD baryogenesis, condensate fragmentation, Q-ball formation and, finally, decay to NLSPs, in order to establish the Q-ball decay temperature $T_{d}$ and resulting NLSP density as a function of the potential parameters.

     Using the semi-analytical method introduced in \cite{fd1}, in which fragmentation is studied by perturbing the negative pressure AD condensate obtained by averaging over oscillations of the AD field, we will first establish the energy, baryonic charge and diameter of the initial fragments. Due to the suppression of the A-terms expected in GMSB models, the energy to (global) charge ratio of the initial fragments is much larger than that of a Q-ball of the same charge. The subsequent evolution of such fragments has been studied numerically for the case of gravity-mediated Q-balls \cite{num}. In that case it was shown that the initial fragments lose energy by producing pairs of positive and negative charge Q-balls (we will refer to these as '$\pm$Q-ball pairs' in the following). No numerical analysis of an equivalent resolution exists as yet for the case of gauge-mediation (although earlier numerical studies do exist which support the formation of $\pm$Q-ball pairs in gauge-mediation \cite{kksim}). However, we can expect a similar dissipation of the energy of the initial fragments via the production of $\pm$Q-ball pairs. In the absence of a full numerical simulation, we will consider different scenarios for the evolution of the initial fragments to a distribution of positive and negative charged Q-balls. To obtain the corresponding Q-ball properties, in particular their decay temperature, we must then match the energy of the initial fragments to the energy of the corresponding final state GMSB Q-balls, following the methods of \cite{fd2}.  

      Given the Q-ball decay temperature $T_{d}$ and the nature of the NLSP, the final NLSP density from Q-ball decay can then be obtained. In order to do this, it is important to consider in detail how Q-balls decay and how the initially relativistic NLSPs from Q-ball decay lose energy and disperse. Q-balls are spatially separated when they decay, therefore if the NLSPs were to rapidly stop via scattering with thermal background particles and annihilate while inhomogeneous, a much lower NLSP density would remain than in the case where the NLSPs can homogenize prior to annihilation. We will show explicitly that the NLSPs from Q-ball decay generally homogenize via relativistic free-streaming well before they annihilate. Therefore the final NLSP density is determined by non-relativistic annihilation of a homogeneous NLSP density produced by Q-ball decay, as is conventionally assumed.

      The NLSPs subsequently decay to gravitinos. If the gravitinos account for the observed dark matter density, they should have a mass less than 1 GeV in order that NLSP decay is compatible with BBN \cite{kazbbn}. In the case where only positively charged Q-balls form in AD baryogenesis, this is a serious obstacle to producing gravitino dark matter via Q-ball decay, since B-conservation combined with R-parity conservation implies that the dark matter particle must satisfy $m_{DM} \approx 2 \GeV$ \cite{fd1,fd2,km1,km2}\footnote{This is also an obstacle to dark matter in the gravity-mediated MSSM. A possible solution in that context was proposed in \cite{rs}, in which the LSP is an axino of mass $\approx$ 2 GeV.}. However, in the case where almost neutral initial fragments break up into positive and negative charged Q-balls, a larger NLSP density will be produced from decay of the $\pm$Q-ball pairs. This is a key feature of the GMSB Q-ball decay model for gravitino dark matter. For the case of stau NLSPs, we will show that the subsequent annihilation of the NLSP density can naturally produce the observed dark matter density with gravitinos in the mass range 0.1-1 GeV.

         Finally, by considering the simplest model for GMSB, based on a vector pair of SU(5) messenger supermultiplets, we will show that the gravitino mass required for dark matter is naturally compatible with the corresponding theoretical gravitino mass from GMSB.

             Our paper is organized as follows. In Section 2 we review the flat-direction potential and the semi-analytical method of \cite{fd1} for determining the nature of the initial AD condensate fragments. We also introduce a simple model for the evolution of these fragments into pairs of positive and negative charged Q-balls. In Section 3 we discuss GMSB Q-ball solutions and compute the Q-ball decay temperature to NLSPs.  We also establish the conditions under which the NLSPs become non-relativistic and homogenize prior to their annihilation. In Section 4 we discuss annihilation of the NLSPs from Q-ball decay and establish the gravitino mass necessary to account for dark matter as a function of the Q-ball decay temperature $T_{d}$. We also discuss the BBN constraints on NLSP decay. In Section 5 we present the results of our global analysis of AD baryogenesis and gravitino dark matter via Q-ball decay. We follow the evolution of flat-direction field from the inflaton-dominated era through the generation of the baryon asymmetry, the fragmentation of the condensate to an ensemble of $\pm$Q-ball pairs and their subsequent decay to a gravitino density via NLSP annihilations and decays. We determine the Q-ball decay temperature and required gravitino dark matter mass as a function of the flat-direction potential parameters. 
In Section 6 we compare the gravitino mass necessary to account for dark matter with the theoretical gravitino mass from GMSB. 
In Section 7 we present and discuss our conclusions.

\section{Flat-direction Potential, Affleck-Dine baryogenesis and condensate fragmentation in GMSB} 

\subsection{GMSB Flat-direction Potential}

   In \cite{fd1} we introduced a phenomenological potential which models the expected behaviour of a GMSB flat direction 
potential as a function of the flat direction field $\Phi$. We will focus on a $d = 6$ $(u^{c}d^{c}d^{c})^{2}$ flat direction in the following. 
The potential is 
\be{e1}  V(\Phi) = m_{s}^2 M_{m}^2 \ln^{2} \left(1 + \frac{|\Phi|}{M_{m}}\right)\left(1 + K \ln \left( \frac{|\Phi|^2}{M_{m}^2} \right) \right) + m_{3/2}^2\left(1 + \hat{K} \ln \left( \frac{|\Phi|^2}{M_{m}^2} \right) \right)|\Phi|^2 
   - cH^2 |\Phi|^2 + 
(A W + h.c.) + 
\left| \frac{\Phi^{5}}{5! \tM^{3}}\right|^2         ~.\ee 
Here
\be{e2}   W = \frac{\Phi^6}{6! \tM^3}     ~,\ee
is the effective superpotential for the flat direction superfield $\Phi$.  
We include the factor $6!$ in \eq{e2} so that the physical strength of the interactions is dimensionally of the order of $\tM$. The scale $\tM$ is usually assumed to be of the order of the Planck scale, however we will consider it to be a free parameter throughout this paper. The log-squared factor in first term in \eq{e1} is due to GMSB with messenger mass $M_{m}$ \cite{gmsbpot}. The factor multiplying this takes into account 1-loop radiative corrections due to gaugino loops once $g |\Phi| \lae M_{m}$, with $K \approx -(0.01-0.1)$ (analogous to gravity-mediated SUSY breaking corrections    
 \cite{km1,km2,km3}). The Standard Model (SM) couplings $g$ at the renormalization scales of interest ($\mu \sim M_{m} \sim 10^{13} \GeV$) are $g \approx 0.6-1$; for simplicity we will set the factor $g$ to 1 in $V(\Phi)$. The second term is due to gravity-mediated SUSY breaking including the 1-loop correction term $ \hat{K}$; for simplicity we will set $\hat{K} = K$.  For the A-term we consider
\be{e3}    A = m_{3/2} +  \frac{a_{o} m_{s}}{\left(1 + 
\frac{|\Phi|^2}{M_{m}^{2}}\right)^{1/2}  }          ~.\ee
The first term in \eq{e3} represents the A-term due to gravity-mediated SUSY breaking. 
The second term models the A-term in gauge-mediated SUSY breaking at $|\Phi| \lae M_{m}$, which is generated at 1-loop from the gaugino masses and is therefore suppressed compared with the A-term in gravity-mediated models. We will consider $a_{o} = \alpha/(4 \pi) \approx 0.01$ to be a typical value, where $\alpha = g^2 /4\pi$ and $g$ is the relevant SM gauge coupling. (We will however consider the effect of varying $a_{o}$.) The suppression factor $(1 + |\Phi|^2/M_{m}^{2})^{-1/2}$ models the $1/|\Phi|$ suppression of the GMSB A-term at $|\Phi| \gg M_{m}$ \cite{gmsbpot}.

\subsection{Affleck-Dine Baryogenesis}

        As $H$ decreases, the $|\Phi| \neq 0$ minimum becomes unstable and the $\Phi$ field begins oscillation in its real and imaginary directions. The B-violating A-terms induce a phase difference between the oscillations in the real and imaginary directions, such that at late times $a^{3/2}\Phi$ (where $a^{3/2}$ takes into account the effect of expansion) describes an ellipse in the complex plane, corresponding to a baryon   
 asymmetry \cite{ad}.
Due to the suppression of the A-terms in GMSB, the condensate is typically highly elliptical (a precessing ellipse, to be precise \cite{fd1}). This means that there is much more energy in the GMSB AD condensate than would be expected if it were entirely made of baryonic charge (corresponding to a circular condensate, so-called "Q-matter"), therefore the condensate is largely neutral with respect to global charge. This fact will play a key role in the Q-ball decay scenario for gravitino DM.

         AD baryogenesis occurs during matter domination by inflaton oscillations. The degree to which the baryon density is subsequently diluted relative to the entropy density depends on the duration of inflaton matter domination, with greater dilution corresponding to a lower reheating temperature, $T_{R}$. $T_{R}$ is chosen to reproduce the observed baryon asymmetry.

\subsection{Initial Fragmentation of the AD Condensate}

       Once the baryon asymmetry is established, perturbations of the AD condensate, corresponding to primordial density perturbations, will grow. The condensate will subsequently fragment when these perturbations are comparable to the mean energy in the condensate. In \cite{fd1} this was studied by following the growth of perturbations due to the average negative pressure in the condensate, which allows the expansion rate at fragmentation, $H_{frag}$, and the wavelength of the dominant perturbation mode, $\lambda_{frag}$, to be determined for a given set of potential parameters. This then determines the energy and baryonic charge of the initial AD condensate lumps.

     We first summarize the method of \cite{fd1}.
The key observation is that the average (negative) pressure in an elliptical condensate is equal to the pressure in a circular condensate, with the constant $|\Phi|$ of the circular condensate set equal to the amplitude of oscillation of the elliptical condensate. In this way the rate of growth of perturbations $\alpha(t)$ of an elliptical condensate can be studied using the same method as for a circular condensate \cite{ks}. This greatly simplifies the analysis of fragmentation as a function of the model parameters, since it allows a semi-analytical approach which enables a wide range of model parameters to be easily scanned. The semi-analytical approach also has the advantage of being robust, in that it is independent of numerical issues such as the  resolution of a lattice simulation. It can therefore complement numerical simulations.

The growth of perturbations of the amplitude and phase satisfies $\delta \varphi, \delta \theta \propto e^{S(t)}$, where $\Phi = \varphi e^{i \theta}/\sqrt{2}$. The rate of growth is then $\alpha = \dot{S}$, where  \cite{fd1,ks}
 \be{e17}  \alpha^2 = \frac{|\bfk|^2}{a^2} \frac{1}{ \left( 
V^{''} + 3 \dot{\Omega}^2 \right) } \left( \dot{\Omega}^2 - 
V^{''} - 16 \frac{|\bfk|^2}{a^2} \frac{ \dot{\Omega}^4 } {\left(V^{''} + 3 \dot{\Omega}^{2}\right)^{2} } \right)     ~.\ee 
Perturbation growth is possible once the physical wavenumber $|\bfk/a|$ of a given perturbation is smaller than a maximum value, $|\bfk_{max}/a|$, where  
\be{e18} \left| \frac{\bfk_{max}}{a} \right|^2 \equiv K_{m}^2 \approx \left( \frac{4 \dot{\Omega}^2}{V^{''} + 3 \dot{\Omega}^2}\right) \times
\left(\dot{\Omega}^2 - V^{''}\right)   ~.\ee
The growth of a perturbation of (comoving) wavenumber ${\bf k}$, once it is smaller than the critical wavenumber for growth, is then given by  
\be{e19} S(t) = \int_{t_{*}}^{t} \alpha(t) dt   ~,\ee
where $t_{*}$ is the time at which the mode begins to grow, corresponding to the time when $|\bfk/a| = K_m$. 
The subsequent growth of the perturbation follows from 
\be{e20}  S(t) = \int_{\alpha_{*}}^{\alpha}  \frac{\alpha(\bfk,a)}{a H} da    ~,\ee   
where $a_{*}$ is the scale factor at $t_{*}$ and 
\be{e21} \alpha(\bfk,a)  
\approx \left( \frac{|\bfk|^2}{a^2} \frac{ \left( \dot{\Omega}^2 - V^{''} \right) }{ \left( V^{''} + 3 \dot{\Omega}^2 \right)}    \right)^{1/2}   ~.\ee
With $H \propto a^{-3/2}$ during inflaton-domination, this gives
\be{e22} S(\bfk,a) = 2 \alpha(\bfk,a)  \left( 1 - \frac{a_{*}^{1/2}}{a^{1/2}} \right) H^{-1} ~.\ee 
At a given value of $a$, the mode with the maximum growth is found by maximizing $S(\bfk,a)$ with respect to $|\bfk|$. With $\alpha(\bfk,a) \propto |\bfk|$ and $ a_{*}(k) = K_{m}^{-1} |\bfk|$, this is maximized at 
\be{e24} \frac{|\bfk|_{max}}{a} = \frac{4}{9} K_{m}  ~.\ee 
The value of $S$ for this mode is 
\be{e24a} S = \frac{2}{3} \alpha H^{-1}     ~.\ee
Here $\alpha$ is determined by the value of $\dot{\Omega}$ and $V$ for the condensate at a given $a$. 
The condition for the condensate to fragment is that
\be{e25} \frac{\delta R}{R}  = \frac{\delta R_{o}}{R_{o}} e^{S} \; \gae 1 \; ~,\ee
for the mode with wavenumber \eq{e24}. $\delta R_{o}/R_{o} \approx \delta \rho/\rho \approx 10^{-4}$ is due to the primordial density perturbation, which also perturbs the energy density of the condensate. The value of $|{\bf k}|_{max}/a$ when this is first satisfied, $|{\bf k}|_{frag}/a$,  will determine the size of the fragments. 
The diameter of the fragments is then given by
\be{e25}   \lambda_{frag} \approx \frac{2 \pi }{\left(\left|\bfk\right|_{frag}/a\right)}    ~.\ee
Their global $U(1)$ charge ($Q = 3B$) is 
\be{e25a}  Q \approx   \lambda_{frag}^3 n_{Q} ~\ee
and the energy of the initial condensate fragment is
\be{e25b}   E \approx  \lambda_{frag}^{3} \rho   ~,\ee 
where we consider each fragment to come from a cube of side $\lambda_{frag}$.

\subsection{Evolution of the initial condensate fragments}

        The initial fragments are almost neutral, corresponding to a highly elliptical condensate. Therefore, the energy to (global) charge ratio, $E/Q$, of the condensate fragments in GMSB models is large compared with the AD scalar mass. As a result, the fragments will lose energy to reach a lower energy state. The scalar field dynamics of this process is highly non-linear and can only be fully understood via a numerical simulation of the evolution of the fields. A high-resolution simulation has been performed for the case of AD baryogenesis in gravity-mediated SUSY breaking models \cite{num}, but as yet there is no equivalent simulation for gauge-mediated models\footnote{Earlier GMSB simulations do exist \cite{kksim}, which support the break-up of the initial fragments into $\pm$Q-ball pairs.}. Nevertheless, since the non-linear dynamics is expected to be similar in the gauge-mediated case (in particular, in the limit where the AD field at initial fragmentation is close to or less than the messenger mass, in which case the potential will be similar to a gravity-mediated potential), we can use the results of \cite{num} as the basis for a simple model for the evolution of the condensate fragments in gauge-mediated models.  

       The results of \cite{num} can be summarized as follows. They consider an AD condensate forming initial fragments with $E/mQ \approx 1/\epsilon$. ($m$ is the AD scalar mass in \cite{num}.) The case of most relevance is a 3D simulation with $\epsilon = 0.01$ (denoted "3D5" in \cite{num}). This is a gravity-mediated flat-direction with $K = -0.1$. In this case the initial condensate fragments from AD baryogenesis (called "first generation Q-balls" in \cite{num}) break up into a distribution of $\pm$Q-ball pairs with $E/m|Q| \approx 1$. 
The charge distribution of the Q-balls (Figure 25 of \cite{num}) shows that the mean charge $|Q|$ of the final Q-balls is 
approximately 4-5 times larger than the mean charge of the initial fragments. The initial fragment energy is $E_{frag} \approx 100 m|Q|_{frag}$, where $|Q|_{frag}$ is the charge of the initial fragment. Therefore with final charge of the Q-balls given by $|Q| \approx (4-5) |Q|_{frag}$ we find $E_{frag} \approx (20-25) m |Q|$. The energy of the final Q-balls is $E_{Q} \approx m |Q|$. Therefore the 
ratio of the initial fragment energy to the final Q-ball energy in the 3D5 simulation of \cite{num} is $E_{frag} \approx (20-25) E_{Q}$.

    Based on this, we will model the decay of the fragments to $\pm$Q-ball pairs by a simple model in which the initial fragments break up into $n_{p}$ pairs of $\pm$Q-balls of equal energy. (The complete fragmentation process is illustrated schematically in Figure 1.) The energy of the Q-balls is then $E_{Q} = E_{frag}/2n_{p}$. The limiting case corresponds to $n_{p} = 1$, with the fragments breaking up into a single pair of $\pm$Q-balls. The simulation of \cite{num} indicates $n_{p} \approx 10-15$ may be more realistic, therefore we will also consider larger $n_{p}$.

\begin{figure}[htbp]
\begin{center}
\epsfig{file=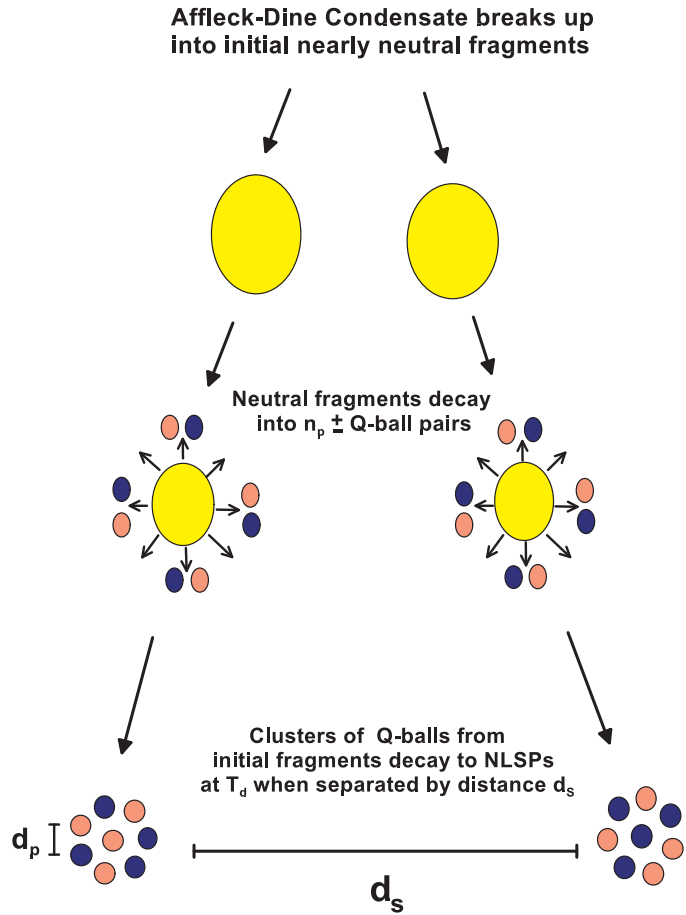, width=0.5\textwidth, angle = -0}
\caption{Schematic evolution of the Affleck-Dine condensate fragments. The nearly neutral initial condensate decay into pairs of positive (light) and negative (dark) Q-balls. The resulting clusters of Q-ball later decay to NLSPs at $T_{d}$ when separated by $d_{s}$. The individual Q-balls in the clusters are separated by $d_{p}$ at $T_{d}$.}
\label{fig1}
\end{center}
\end{figure}

\section{GMSB Q-ball solutions and Q-ball decay}

  The squarks which form the Q-ball decay primarily to quarks and gauginos, 
 with the gauginos subsequently decaying to NLSPs. (If decay to gauginos is kinematically excluded then the Q-balls decay directly to quarks and NLSPs.) 
The Q-ball decay rate to fermions has been calculated in \cite{cole2}. In the saturated case, which occurs when $g \varphi_{o} \gg \omega$, where $\varphi_{o}$ is the magnitude of the field inside the Q-ball and $\omega$ is the effective mass of the Q-ball squarks, the decay rate of the Q-ball is given by \cite{cole2}
\be{e26} \Gamma_{d} = \frac{\omega^{3} R^{2}}{48 \pi Q}   ~,\ee 
where $R$ is the radius of the Q-ball. The Q-balls decay once $\Gamma_{d} \approx H$, with the expansion rate given by $H = k_{T}T^{2}/M_{p}$, where $k_{T} = (\pi^2 g(T)/90)^{1/2}$ and $g(T)$ is the effective number of massless degrees of freedom. Therefore the Q-ball decay temperature is  
\be{e12} T_{d} \approx  \left( \frac{\omega^{3} R^{2} M_{P}}{48 \pi k_{T_{d}} Q} \right)^{1/2}     ~.\ee

      In order to estimate the Q-ball decay temperature, we therefore need to obtain the Q-ball parameters $\omega$ and $R$ and the Q-ball charge $Q$ of the $\pm$Q-balls. 

Our method is to find GMSB Q-ball solutions which match the Q-ball energy from the fragmentation model, $E_{frag}/2n_{p}$, for a given $n_{p}$. 
The GMSB Q-balls are minimum energy solutions 
for a fixed global charge $Q$. The potential is given by \eq{e1}, but neglecting the A-term, $H^2$ mass term and non-renormalizable term, which are negligibly small inside the Q-ball. The Q-ball solutions have the form 
\be{e6} \Phi = \frac{\varphi(r)}{\sqrt{2}}e^{i \omega t} ~,\ee
where the Q-ball profile $\varphi(r)$ is given by the solution of 
\be{e7}  \frac{\partial^2 \varphi}{\partial r^2} + \frac{2}{r}\frac{\partial \varphi}{\partial r} = \frac{\partial V}{\partial \varphi} - \omega^2 \varphi  ~,\ee
with boundary conditions $\varphi^{'}(r) = 0 $ as $r \rightarrow 0$ and $\vphi(r) \rightarrow 0$ as $r \rightarrow \infty$.
The total charge and energy of the Q-ball are then 
\be{e8} Q = \int^{\infty}_{0} 4\pi r^2 \omega \varphi(r)^2 dr  \;\;\;;\;\;\; E = \int^{\infty}_{0}  4 \pi r^2 \left[ \frac{1}{2} \left(\frac{\partial \varphi}{\partial r}\right)^2 
+ V(\varphi) + \frac{\omega^2 \varphi^2}{2} \right] dr. ~\ee 
For each value $\omega$ there is a unique Q-ball solution with energy $E$ and charge $Q$. By matching the energy $E$ from the fragmentation model to the corresponding Q-ball solution with the same energy, 
the values of $\omega$, $Q$, $E/Q$ and $R$ can be obtained\footnote{Since in general the Q-ball profiles are not exactly step-like, as assumed in \cite{cole2}, we will define $R$ in our solutions to be the radius within which $90\%$ of the total energy is found.}.  
     
    The analysis of \cite{cole2} is for a thin-walled Q-ball of radius $R$, with a step-like profile for $\varphi(r)$. In reality, Q-balls in the MSSM are not generally thin-walled. However, in the limit where the field in the Q-ball significantly larger than the messenger mass, the Q-ball has a broad profile \cite{fd2} and a thin-walled approximation is a reasonable approximation to the Q-ball. In the opposite limit, where the field is small compared with the messenger mass, the Q-balls will be similar to gravity-mediated type Q-balls, which are thick-walled with a Gaussian-like profile. A very recent analysis \cite{qqnew} has considered Q-ball decay with a general profile. It is shown that, in the saturated limit, the decay rate is similar to \eq{e26} for the case of GMSB Q-balls with a large field strength compared to the messenger mass and somewhat larger that \eq{e26} (typically by a factor $\sim 20$) for the case of gravity-mediated-like Q-balls. In the latter case the decay temperature can be a factor $\sim 4-5$ larger than  \eq{e12}. In fact, we will show that the Q-balls with field strength large compared with the messenger mass are the ones primarily responsible for gravitino dark matter which is consistent with BBN, therefore \eq{e26} should be a good approximation to the Q-ball decay rate.

\section{NLSPs and Gravitino Dark Matter from Q-ball Decay}

\subsection{Q-ball decay to NLSPs}

          Provided that the mass of the NLSP, $m_{\chi}$, is less than $E/Q$, decay to NLSPs is kinematically possible. The NLSPs will initially be produced with relativistic or near-relativistic energy $O(E/Q)$. The final NLSP density will then depend on the subsequent evolution of the initially relativistic NLSPs. 

         The conventional picture of Q-ball decay assumes that the NLSPs homogenize and become non-relativistic prior to annihilation \cite{kmdm,yama}. It is important to check this assumption. The NLSPs from Q-ball decay will lose energy via scattering with relativistic particles in the thermal background. They will then slow and become non-relativistic within a distance $d_{NR}$. At Q-ball decay, the Q-balls from different initial fragments will be in clusters separated by a distance $d_{S}$ (Figure 1). Therefore if $d_{NR} \gg d_{S}$, the NLSPs will cross between the Q-ball clusters via relativistic free-streaming before they slow by scattering and become non-relativistic. In this case the NLSPs will become homogeneous and non-relativistic before they annihilate. We discuss the calculation of $d_{NR}$ and $d_{S}$ in the Appendix. We will show in the next section that  $d_{NR} \gg d_{S}$ is easily satisfied for the GMSB Q-balls of interest to us here. We can  therefore compute the final NLSP density under the assumption of a homogeneous and non-relativistic NLSP density from Q-ball decay.

       We next consider the evolution of the NLSP density from Q-ball decay. We first consider a density of neutral NLSPs. Q-balls decay away completely over a time $\approx H^{-1}(T_{d})$. From \eq{e26} the Q-balls decay according to
\be{e10} \frac{d Q}{d t} = -K_{d}  \;\;\;;\;\;\; K_{d} \equiv  \frac{\omega^{3} R^{2}}{48 \pi}   ~.\ee
Therefore the homogeneous non-relativistic NLSP density $n$ will increase at a constant rate as the Q-balls decay and inject NLSPs. Eventually the annihilation rate of the NLSP density, $\Gamma_{ann} = n <\sigma v>$, where $\sigma$ is the annihilation rate and $v$ the relative velocity of the NLSPs, and the average is over the NLSP distribution, will become equal to the expansion rate, at which point annihilations become effective and the density no longer increases. The final density of non-thermal NLSPs from Q-ball decay is therefore
\be{f11} n_{ann}(T_{d}) \approx \frac{H(T_{d})}{<\sigma v>}    ~.\ee
This density will be achieved provided that the total NLSP density from Q-ball decay in the absence of annihilations, $n_{TOT}(T_{d})$, is larger than $n_{ann}(T_{d})$. This will be shown to be generally true in the cases of interest to us here. 

    In the case where the NLSPs carry a gauge or global charge, as in the case of stau NLSPs, the Q-balls will produce equal number densities of NLSPs and anti-NLSPs. In this case $n_{ann}(T_{d})$ in \eq{f11} gives the number density of the NLSPs (which is equal to the number density of anti-NLSPs), the total NLSP number density is the sum of NLSP and anti-NLSP number densities and $<\sigma v>$ in \eq{f11} is the total annihilation cross-section of an NLSP with both NLSPs and anti-NLSPs from Q-ball decay.

\subsection{Dark matter gravitino mass}

    Given the NLSP density from Q-ball decay, we can determine the value of the gravitino mass $m_{3/2}$ necessary to account for the observed dark matter density. By R-parity conservation, one gravitino is produced per NLSP decay, therefore $n_{3/2} = n_{NLSP}$. The gravitino density at present is then
\be{f12} n_{3/2}(T_{\gamma}) = \frac{g(T_{\gamma})}{g(T_{d})}\frac{T_{\gamma}^{3}}{T_{d}^{3}} n_{ann}(T_{d})    ~.\ee
Therefore gravitino mass required to account for a dark matter abundance $\Omega_{3/2}$ is $m_{3/2} = \rho_{c} \Omega_{3/2}/n_{3/2}$, where $\rho_{c}$ is the critical density.

    We next apply this to the case of stau NLSPs, which we will focus on in the following. There are two dominant annihilation channels for a stau \cite{stauann}: (i) 
$\stau \stau^{*} \rightarrow \gamma \gamma,\; \gamma {\rm Z}$(where $\stau^{*}$ denotes the anti-stau), and (ii) $\stau \stau \rightarrow \tau \tau$. The average cross-sections times relative velocity for these processes are \cite{stauann} 
\be{f10} <\sigma v>_{\stau \stau^{*}} \approx \frac{4 \pi \alpha_{em}^{2}}{m_{\stau}^{2}}     ~\ee 
and
\be{f10a} <\sigma v>_{\stau \stau} \approx \frac{16 \pi \alpha_{em}^{2} m_{\tilde{B}}^{2}}{\cos^{4}\theta_{w} \left(m_{\stau}^{2} + m_{\tilde{B}}^{2} \right)^2}     ~,\ee
where $\tilde{B}$ is the bino.  
The stau annihilation rate is then 
\be{f10b} \Gamma_{ann} = n_{\stau} <\sigma v>_{\stau \stau} 
+ n_{\stau^{*}} <\sigma v>_{\stau \stau^{*}} ~.\ee 
Therefore, using $n_{\stau} = n_{\stau^{*}}$, we have
\be{f10c} \Gamma_{ann} = n_{\stau} <\sigma v> \;\;\; ; \;\;\; 
<\sigma v> \equiv   <\sigma v>_{\stau \stau}  +
 <\sigma v>_{\stau \stau^{*}}    ~.\ee
The number of staus from Q-ball decay is then
\be{f11a} n_{\stau} \approx \frac{H(T_{d})}{<\sigma v>}   ~.\ee     
The same number of $\stau^{*}$ are produced. The total NLSP density is therefore 
\be{f11b}  n_{NLSP}(T_{d}) \equiv n_{\stau} + n_{\stau^{*}} \approx 
\frac{2 H(T_{d})}{<\sigma v>}   ~.\ee
From \eq{f10} and \eq{f10a}, we see that $<\sigma v>_{\stau \stau} \; \ll \; <\sigma v>_{\stau \stau^{*}}$ in the limit where 
$m_{\tilde{B}} \gg m_{\stau}$, while
$<\sigma v>_{\stau \stau}$ and  $<\sigma v>_{\stau \stau^{*}}$ 
become comparable at $m_{\tilde{B}} \approx m_{\stau}$. In the following will assume $<\sigma v>_{\stau \stau} = <\sigma v>_{\stau \stau^{*}}$, noting that $<\sigma v>$ could be smaller by a factor of 2 if the bino is sufficiently heavy. 

Using $<\sigma v> = 2 <\sigma v>_{\stau \stau^{*}}$, we find that the gravitino mass required to account for the observed density of dark matter in the case of stau NLSPs is 
\be{f14a} m_{3/2} \approx \frac{g(T_{d})}{g(T_{\gamma})}
\frac{T_{d} \Omega_{3/2} \rho_{c}}{k_{T_{d}} T_{\gamma}^{3} } \frac{  4 \pi \alpha_{em}^{2} M_{p}}{m_{\stau}^{2}}    ~.\ee
Therefore 
\be{f14} m_{3/2} \approx 0.06 \GeV \times 
 \left( \frac{g(T_{d})}{10.75}\right)^{1/2} 
\left( \frac{T_{d}}{10 \MeV} \right)
\left( \frac{1 \TeV}{m_{\stau}} \right)^{2}     ~.\ee
In this have used $g(T_{d}) = 10.75$ for $T_{d} \sim 10 \MeV$, $\alpha_{em} = 1/128$, $\Omega_{3/2}h^2 = 0.1126$ \cite{wmap7}  and  $\rho_{c} = 8.1 h^2 \times 10^{-47} \GeV^4$.

\subsection{BBN Constraints}

 The BBN constraints on NLSP decay to gravitinos have been analyzed in detail in \cite{kazbbn}. The constraints follow from the decay of NLSPs at $T \lae T_{BBN} \approx 1 \MeV$, which can modify light element abundances formed during BBN. The results of \cite{kazbbn} can be summarized as follows. Non-thermal stau or sneutrino NLSPs can produce gravitino dark matter and remain consistent with BBN if the NLSP mass is greater than around 300 GeV and $m_{3/2} \lae 1 \GeV$ (Figs.14 and 16 of \cite{kazbbn}), while a bino NLSP of mass greater than about 300 GeV is consistent with BBN if $m_{3/2} \lae 2 \times 10^{-2} \GeV$ (Figs. 9 and 10 of \cite{kazbbn}). There is a large hadronic component in the decay products of bino decay, which leads to stronger constraints than the case of a stau or sneutrino NLSP, which have primarily 
radiative decays with only a small hadronic component. (The stau can also catalyze formation of $\;^{6}$Li via formation of $^{4}$He-$\tilde{\tau}$ bound states.) The analysis of \cite{kazbbn} considers only two NLSP masses, 100 GeV and 300 GeV. However, the upper bound is generally expected to become weaker for larger NLSP mass, due to the earlier time of decay. 

  The upper bound on the gravitino mass can be understood as a lower bound on the time of NLSP decay. The NLSP decay rate is 
proportional to $m_{NSP}^5/m_{3/2}^2$. Therefore, for a given NLSP mass, the NLSPs decay earlier as the gravitino mass becomes smaller. If the NLSP has primarily radiative decays, as in the case of stau and sneutrino NLSPs, then as long as the decay occurs sufficiently before the formation of light element abundances at $T \approx 0.08 \MeV$, the electromagnetic cascade from NLSP decay will not dissociate the light elements. 

      Therefore, in order to evade BBN constraints on stau decay to gravitinos, the dark matter gravitino mass must be less than 1 GeV. From \eq{f14} we see that this 
requires that the Q-balls decay at a temperature of the order of 10 MeV.

\section{Global Analysis of Affleck-Dine Condensate Fragmentation, Q-ball Decay and Gravitino Dark Matter}

   In this section we will present the results of our global analysis of AD baryogenesis, Q-ball formation and decay to gravitino dark matter, using the methods discussed in the previous sections. We will focus on the case of stau NLSPs.

    There are a number of conditions that must be satisfied in order to have successful production of gravitino dark matter via Q-ball decay to stau NLSPs: (i) nucleosynthesis requires that $T_{d} > 1 \MeV$, in order that the baryon asymmetry is released from the Q-balls before the onset of nucleosynthesis at $T_{BBN} \approx 1 \MeV$; 
(ii) the gravitino mass should be less than 1 GeV, in order that stau decay does not alter element abundances from BBN; (iii) the Q-balls should be able to decay directly to stau NLSPs. This requires that $E/Q > m_{\stau}$. The present LHC CMS lower bound on the stau mass is $m_{\stau} > 223 \GeV$ \cite{lhc}. We will conservatively set $m_{\stau} = 250 \GeV$ throughout; larger values of $m_{\stau}$ will give lower values of $m_{3/2}$.

      In Table 1 we present the Q-ball decay temperature and dark matter gravitino mass as a function of $m_{s}$ for the case $n_{p} = 10$, $a_{o} = 0.01$, $\tM = M_{p}$ and $M_{m} = 10^{13} \GeV$. (We set $K = -0.01$ throughout; our main results are not sensitive to the value of $K$, which only becomes important for gravity-mediated type Q-balls.) We consider large values for the AD squark mass, $m_{s} \geq 1 \TeV$, as suggested by the non-observation of squarks at the LHC.  We find that dark matter gravitino production satisfies all BBN and experimental constraints in all the examples shown in Table 1. 
This is our main result: gravitino dark matter can successfully originate from Q-ball decay to NLSPs in GMSB. 

The gravitino mass for the cases in Table 1 is in the range 0.1-1 GeV, while the Q-ball decay temperature is in the range 4.4-7.7 MeV. 
(In the case where the bino mass is much larger than the stau mass, $m_{3/2}$ can be smaller by a factor of about 2 for a given $T_{d}$.) The value of $E/Q$ is in the range 400-706 GeV. The Q-balls generally have $\varphi(0)/M_{m} > 10$, therefore they are gauge-mediated type Q-balls, dominated by the logarithmic plateau of the potential.

   In Table 2 we show the corresponding $d = 6$ GMSB AD baryogenesis and condensate fragmentation parameters. The reheating temperature is chosen to reproduce the observed baryon asymmetry and is in the range 44-87 GeV.

    In Table 3 we give the values of $d_{NR}$ and $d_{S}$ for the Q-balls in Table 1. We find that $d_{NR} \gg d_{S}$, with $d_{NR} \sim 10^{18} \GeV^{-1}$ and $d_{S} \sim 10^{13} \GeV^{-1}$. Therefore the stau NLSPs from the decay of these Q-balls will homogenize by free-streaming and become non-relativistic well before annihilating. This is generally true of all the cases we consider here.

    In Tables 4 to 7 we show how these results depend on the other parameters of the model. We show the effect of varying $\tM$ in Table 4, the effect of varying $M_{m}$ in Table 5, the effect of varying $a_{o}$ in Table 6 and the effect of varying $n_{p}$ in Table 7. All tables use the first entry of 
Table 1 as their starting point.   

    From Table 4 we find that larger values of $\tilde{M}$ are preferred. Decreasing $\tilde{M}$ from $M_{p}$ causes $m_{3/2}$ to rapidly increase above 1 GeV. Thus the value of $\tilde{M}$ which is often considered the most natural value, $\tilde{M} \sim M_{p}$, is also preferred by gravitino dark matter from Q-ball decay.  

   From Table 5 we find that decreasing the messenger mass $M_{m}$ causes both $T_{d}$ and the gravitino mass to decrease. There is a small window from  $M_{m} = 8 \times 10^{12} \GeV$ to $1 \times 10^{13} \GeV$ for which $T_{d}$ is greater than 1 MeV while $m_{3/2}$ is less than 1 GeV. Larger $M_{m}$ causes the gravitino mass to violate the BBN bound while smaller $M_{m}$ causes $T_{d}$ to violate the BBN bound. 
The gravitino mass can be decreased significantly by considering either a larger bino mass or a stau mass larger than 250 GeV. For example, increasing the stau mass to 500 GeV would decrease the gravitino mass for the case $M_{m} = 2 \times 10^{13} \GeV$ in Table 5 to 0.63 GeV, so increasing the upper bound on the allowed range of $M_{m}$ values.

   From Tables 4 and 5 and Figure 2  we find that there is a strong correlation between the value of $\varphi(0)/M_{m}$, where $\varphi(0)$ is the value of the field at the centre of the Q-ball, and the dark matter gravitino mass. The Q-ball decay temperature and gravitino mass rapidly increase as $\varphi(0)/M_m$ approaches 1. This is due to the transition of the Q-balls from gauge-mediated to gravity-mediated type. This is confirmed by the fact that $\omega \rightarrow m_{s}$ in the limit
where $\varphi(0)/M_m \ll 1$, as expected for gravity-mediated type Q-balls. We also find the $\varphi(0)/M_m$ and the gravitino mass become very sensitive to the model parameters once $\varphi(0)/M_m$ is significantly larger than 1. This is due to the insensitivity of the potential to $\varphi$ at large $\varphi/M_m$, where the potential has only a log-squared dependence on $\varphi$. As a result, small changes in the potential energy can produce large changes in the value of the field.

    From Table 6 we find that the Q-ball decay temperature and $T_{d}$ is essentially independent of the GMSB A-term $a_{o}$. At first sight this seems surprising. However, in GMSB models the baryon asymmetry contributes only a small part of the charge of the $\pm$ Q-balls coming from the almost neutral initial fragments. Therefore the $a_{o}$ term has very little effect on the Q-balls, provided that the AD condensate is strongly elliptical. In this case the only effect of the $a_{o}$ term is to alter the net baryon asymmetry in the AD condensate and so the reheating temperature, however this plays no role in Q-ball decay.  

 From Table 7 we find that decreasing $n_{p}$, corresponding to production of larger charge $\pm$Q-ball pairs from the initial condensate fragments, decreases both $T_{d}$ and $m_{3/2}$. Thus smaller $n_{p}$ is preferred by the 
gravitino BBN bound,  provided that the Q-ball decay temperature is greater than 1 MeV.

    In obtaining these results we have assumed that a sufficient number density of NLSPs is produced from Q-ball decay to be able to reach the annihilation density $n_{NLSP}(T_{d})$ in \eq{f11b}. This requires that 
$n_{TOT} >  n_{NLSP}(T_{d})$, where $n_{TOT}$ is the total NLSP number density from complete Q-ball decay in the absence of annihilations. $n_{TOT}$ is given by 
\be{xt1} n_{TOT} \approx \frac{2 n_{p} Q}{d_{S}^{3}}   ~.\ee 
This follows because the Q-ball clusters are separated by $d_{S}$ at decay and contain $2n_{p}$ Q-balls of charge $\pm$Q, with one NLSP being produced per unit Q-ball charge. 

   For the first entry in Tables 1-3, we have $n_{p} = 10$, $d_{S} = 1.4 \times 10^{13} \GeV^{-1}$ and $Q = 5.6 \times 10^{24}$, which gives $n_{TOT} = 4.1 \times 10^{-14} \GeV^3$. With $T_{d} = 4.4 \MeV$, $k_{T_{d}} \approx 1$ and $m_{\stau} = 250 \GeV$, \eq{f11b} gives $n_{NLSP}(T_{d}) = 6.5 \times 10^{-16} \GeV^3$. Therefore $n_{TOT} \gg n_{NLSP}(T_{d})$ and so the NLSPs will annihilate. 
This also makes clear the importance of annihilations in limiting the NLSP density from Q-ball decay. In the absence of annihilations, the NLSP density would be $\sim 10^2$ times larger, requiring a very much smaller gravitino mass. This would not be compatible with the large messenger mass necessary to keep $E/Q$ large enough to allow decay to NLSPs.

\begin{table}[h]
\begin{center}
\begin{tabular}{|c|c|c|c|c|c|c|c|c|c|}
 
\hline $m_s (\TeV)$	&	$\tilde{M} (\GeV)$  & $M_{m} (\GeV)$  & $\varphi(0)/M_{m}$	&	$	\omega^2/m_{s}^{2} $   & $E (\GeV)$   & $Q$	&	$E/Q (\GeV)$  &  $T_{d} (\GeV)$ &	$m_{3/2} (\GeV) $                      \\
\hline	$1.0 $	&	$2.43 E 18$	&	$1 E 13$   &   $15.2$ 	&	$	0.108$		&	$2.23 E27$ 	&  $5.59 E24$	&	$399$ & $0.0044$	& $0.40$	\\
\hline	$1.5 $	&	$2.43 E 18$	&	$1 E 13$   &   $17.4$	&	$0.093$			&	$1.97 E27$ 	&  $6.45 E25$		&	$557$ & $0.0067$	& $0.55$	\\
\hline	$2.0 $	&	$2.43 E 18$	&	$1 E 13$   &   $18.9$ 	&	$0.083$			&	$2.07 E27$ 		&  $2.79 E24$	&	$706$ & $0.0077$	& $0.67$	\\

\hline     
 \end{tabular} 
 \caption{\footnotesize{Q-ball properties and dark matter gravitino mass for $n_{p} = 10$ and $a_{o} = 0.01$ as a function of varying $m_{s}$.}}  
 \end{center}
 \end{table}

\begin{table}[h]
\begin{center}
\begin{tabular}{|c|c|c|c|c|c|c|c|c|}
 
\hline $m_s (\TeV)$	&	$\tilde{M} (\GeV)$  & $M_{m} (\GeV)$  &   $T_{R} (\GeV)$	& $H_{frag} (\GeV)$   & $\lambda_{frag} (\GeV^{-1})$    &  $|\Phi|_{frag}/M_{m}$  &  $E_{frag} (\GeV)$	&	$Q_{frag}$   \\
\hline	$1.0 $	&	$2.43 E 18$	&	$1 E 13$   &   $44.0$  &	    $4.27$		&	$0.050$ 		&  $2.60$		&	$4.46 E28$ & $1.17 E23$	\\
\hline	$1.5 $	&	$2.43 E 18$	&	$1 E 13$   &   $66.1$	&			$6.07$		&	$0.035$ 		&  $2.85$		&	$4.17 E28$ & $5.43 E22$\\
\hline	$2.0 $	&	$2.43 E 18$	&	$1 E 13$   &   $86.9$ 	&			$7.69$		&	$0.027$ 		&  $3.04$ 		&	$3.97 E28$ & $3.13 E22$	\\

\hline     
 \end{tabular} 
 \caption{\footnotesize{AD baryogenesis and initial condensate fragmentation parameters for $n_{p} = 10$ and $a_{o} = 0.01$ as a function of varying $m_{s}$.}}  
 \end{center}
 \end{table}

\begin{table}[h]
\begin{center}
\begin{tabular}{|c|c|c|c|c|c|c|c|}
 
\hline $m_s (\TeV)$	&	$\tilde{M} (\GeV)$  & $M_{m} (\GeV)$  & $d_s (\GeV^{-1})$	& $d_{NR}	(\GeV^{-1})$        \\
\hline	$1.0 $	&	$2.43 E 18$	&	$1 E 13$   &	$1.46 E13$	&	$2.98 E18$	\\   	
\hline	$1.5 $	&	$2.43 E 18$	&	$1 E 13$   &	$8.53 E12$	&	$9.62 E17$	\\   		
\hline	$2.0 $	&	$2.43 E 18$	&	$1 E 13$   &	$5.77 E12$	&	$4.31 E17$	\\

\hline     
 \end{tabular} 
 \caption{\footnotesize{Q-ball decay parameters for $n_{p} = 10$ and $a_{o} = 0.01$ as a function of varying $m_{s}$. }}  
 \end{center}
 \end{table}

\begin{figure}[htbp]
\begin{center}
\epsfig{file=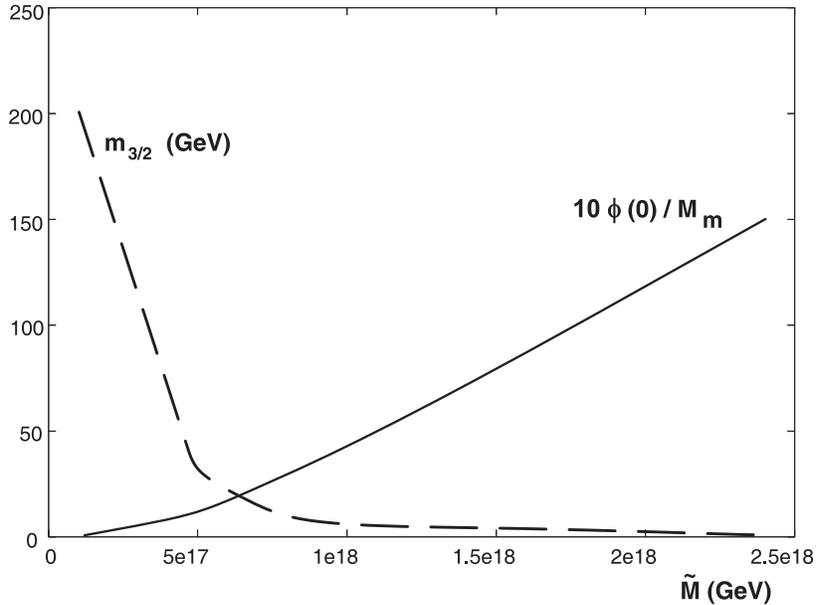, width=0.45\textwidth, angle = -90}
\caption{Plot of $m_{3/2}$ (dashed line, GeV units) and $10\varphi(0)/M_m$ (solid line) versus $\tilde{M}$ for the case $M_{m} = 10^{13} \GeV$.
As $\varphi(0)/M_{m}$ approaches 1, corresponding to the transition from gauge- to gravity-mediated type Q-balls, the dark matter gravitino mass rapidly increases.}
\label{fig1}
\end{center}
\end{figure}

\begin{table}[h]
\begin{center}
\begin{tabular}{|c|c|c|c|c|c|c|c|c|c|}
 
\hline $m_s (\TeV)$	&	$\tilde{M} (\GeV)$  & $M_{m} (\GeV)$  & $\varphi(0)/M_{m}$	&	$\omega^2/m_{s}^2$   & $E (\GeV)$   & $Q$	&	$E/Q (\GeV)$  &  $T_{d} (\GeV)$ &	$m_{3/2}  (\GeV)$                      \\
\hline	$1.0 $	&	$2.43 E 18$	&	$1 E 13$   &   $15.2$ 	&	$	0.108$		&	$2.23 E27$ 	&  $5.59 E24$	&	$399$ & $0.0044$	& $0.40$	\\
\hline	$1.0 $	&	$1 E 18$	&	$1 E 13$   &   $4.5$	&	$0.334$			&	$1.43 E26$ 	&  $2.13 E23$		&	$670$ & $0.052$	& $4.82$	\\
\hline	$1.0 $	&	$8 E 17$	&	$1 E 13$   &   $3.2$ 	&	$0.423$			&	$7.18 E25$ 		&  $9.66 E22$	&	$743$ & $0.095$	& $8.54$	\\
\hline	$1.0 $	&	$5 E 17$	&	$1 E 13$   &   $1.4$ 	&	$0.651$			&	$1.65 E25$ 		&  $1.87 E22$	&	$885$ & $0.30$	& $26.93$	\\
\hline	$1.0 $	&	$2 E 17$	&	$1 E 13$   &   $0.1$ 	&	$1.036$			&	$6.92 E23$ 		&  $6.69 E20$	&	$1033$ & $2.21$	& $201.14$	\\

\hline     
 \end{tabular} 
 \caption{\footnotesize{Q-ball properties and dark matter gravitino mass for $n_{p} = 10$ and $a_{o} = 0.01$ as a function of varying $\tM$.}}  
 \end{center}
 \end{table}

\begin{table}[h]
\begin{center}
\begin{tabular}{|c|c|c|c|c|c|c|c|c|c|}
 
\hline $m_s (\TeV)$	&	$\tilde{M} (\GeV)$  & $M_{m} (\GeV)$  & $\varphi(0)/M_{m}$	&	$	\omega^2/m_{s}^{2}$   & $E (\GeV)$   & $Q$	&	$E/Q (\GeV)$  &  $T_{d} (\GeV)$ &	$m_{3/2}  (\GeV)$    \\
\hline	$1.0 $	&	$2.43 E 18$	&	$5 E 13$   &   $0.3$	&		$0.940$			&	$3.89 E25$ 	&  $5.88 E22$		&	$1004$ & $0.24$	& $20.14$	\\
\hline	$1.0 $	&	$2.43 E 18$	&	$2 E 13$   &   $4.4$	&		$0.342$			&	$5.33 E26$ 	&  $7.87 E23$		&	$667$ & $0.031$	& $2.55$	\\
\hline	$1.0 $	&	$2.43 E 18$	&	$1.5 E 13$  &  $7.4$	&		$0.223$			&	$9.23 E26$ 	&  $1.65 E24$		&	$558$ & $0.018$	& $1.30$	\\
\hline	$1.0 $	&	$2.43 E 18$	&	$1 E 13$   &   $15.2$ 	&	$0.108$			&	$2.23 E27$ 	&  $5.59 E24$		&	$399$ & $0.0044$	& $0.40$	\\
\hline	$1.0 $	&	$2.43 E 18$	&	$9 E 12$   &   $17.6$	&	$0.092$			&	$2.57 E27$ 	&  $6.93 E24$		&	$370$ & $0.0035$	& $0.32$	\\
\hline	$1.0 $	&	$2.43 E 18$	&	$8 E 12$   &   $25.8$	&	$0.059$			&	$5.31 E27$ 	&  $1.78 E25$		&	$298$ & $0.0016$	& $0.15$	\\
\hline	$1.0 $	&	$2.43 E 18$	&	$5 E 12$   &   $33.4$ 	&	$0.042$			&	$1.41 E28$ 	&  $4.01 E25$		&	$255$ & $0.0008$	& $0.076$	\\
\hline	$1.0 $	&	$2.43 E 18$	&	$1 E 12$   &   $502.4$	&	$0.0008$			&	$2.45 E29$ 	&  $6.71 E27$		&	$36$  & $3.49 \times 10^{-6}$	& $0.00031$	\\

\hline     
 \end{tabular} 
 \caption{\footnotesize{Q-ball properties and dark matter gravitino mass for $n_{p} = 10$ and $a_{o} = 0.01$ as a function of varying $M_{m}$.}}  
 \end{center}
 \end{table}

\begin{table}[h]
\begin{center}
\begin{tabular}{|c|c|c|c|c|c|c|c|}
 
\hline $a_{o}$  & $\varphi(0)/M_{m} $	&	$	\omega^2/m_{s}^{2}$   & $E (\GeV)$   & $Q$	&	$E/Q$  &  $T_{d} (\GeV)$ &	$m_{3/2} (\GeV)$                      \\
\hline	$0.1$   &   $15.22$ 	&	$0.108$			& $2.23 E27$ 	&  $5.59 E24$	&	$398.87$ & $0.0044$	& $0.40$	\\
\hline	$0.01$   &   $15.2$ 	&	$0.108$		&	$2.23 E27$ 	&  $5.59 E24$	&	$399.26$ & $0.0044$	& $0.40$	\\
\hline	$0.001$   &   $15.19$ 	&	$0.108$		&	$2.23 E27$ 	&  $5.59 E24$	&	$399.51$ & $0.0045$	& $0.41$	\\
\hline     
 \end{tabular} 
 \caption{\footnotesize{Q-ball properties and dark matter gravitino mass for $n_{p} = 10$ as a function of varying $a_{o}$. Here $m_{s} = 1 \TeV$, 
$M_{m} = 10^{13} \GeV$ and $\tM = 2.43 \times 10^{18} \GeV$.}}  
 \end{center}
 \end{table}

\begin{table}[h]
\begin{center}
\begin{tabular}{|c|c|c|c|c|c|c|c|}
\hline $n_{p}$  & $\varphi(0)/M_{m}$	&	$	\omega^2/m_{s}^{2} $   & $E (\GeV)$   & $Q$	&	$E/Q (\GeV) $  &  $T_{d} (\GeV)$ &	$m_{3/2}   (\GeV)$                      \\
\hline	$1$   &   $38.0$ 	&	$0.036$		&	$2.23 E28$ 	&  $9.43 E25$	&	$236$ & $4.8 E-4$	& $0.043$	\\
\hline	$5$   &   $20.2$ 	&	$0.079$			&	$4.46 E27$ 		&  $1.30 E25$	&	$343$ & $2.4 E-3$	& $0.21$	\\
\hline	$10$   &   $15.2$ 	&	$0.108$		&	$2.23 E27$ 	&  $5.59 E24$	&	$399$ & $4.4 E-3$	& $0.40$	\\
\hline	$25$   &   $10.4$ 	&	$0.162$			&	$8.91 E26$ 		&  $1.85 E24$	&	$482$ & $1.1 E-2$	& $0.95$	\\

\hline     
 \end{tabular} 
 \caption{\footnotesize{Q-ball properties and dark matter gravitino mass as a function of varying $n_{p}$. Here $m_{s} = 1 \TeV$, 
$M_{m} = 10^{13} \GeV$, $\tM = 2.43 \times 10^{18} \GeV$ and $a_{o} = 0.01$.}}  
 \end{center}
 \end{table}

\section{Comparison of the gravitino dark matter mass with the theoretical GMSB gravitino mass}

       In the previous section we showed that it is possible for Q-ball decay in GMSB models to produce non-thermal NLSPs which have the right density to account gravitino dark matter of mass less than 1 GeV. However, a given GMSB model will also predict the gravitino mass. 
We therefore must compare the dark matter gravitino mass with the theoretical gravitino mass. 

        We will consider a standard GMSB framework with $n_5$ vector pairs of messenger chiral supermultiplets $f + \overline{f}$, each transforming as a  ${\bf 5} +  \overline{{\bf 5}}$ of SU(5) \cite{gmsb1}. The superpotential coupling of the messenger supermultiplets to the SUSY breaking supermultiplet is 
\be{c1}   W = \kappa S \overline{f} f    ~.\ee 
The $S$ supermultiplet has expectation value $S = <S> + \theta^2 <F_{S}>$. In this case the masses of the $f$ and $\overline{f}$ scalars and fermions are 
\be{c2}  m_{scalar}^2 = M_{m}^{2} \pm \Lambda M_{m}    ~\ee
and 
\be{c3}  m_{fermion} = M_{m}    ~,\ee 
where $\Lambda = <F_{S}>/<S>$ and $M_{m} = \kappa <S>$. 
This leads to gaugino masses for MSSM gauginos, $\lambda_{i}$, at 1-loop 
\be{c4} m_{\lambda_{i}} = n_{5} \Lambda \frac{\alpha_{i}}{4 \pi}     ~\ee
and to scalar masses at 2-loops
\be{c5}  m_{s}^{2} = 2 n_{5} \Lambda^{2} \sum_{i=1}^{3} C_{i} \left(\frac{\alpha_{i}}{4 \pi} \right)^{2}    ~.\ee
Here $C_{i}$ are the Casimir coefficients for the scalar multipet of interest, with $i = 1,2,3$ for the $U(1)_{Y}$, $SU(2)_{L}$ and $SU(3)_{c}$ gauge groups. 
The gravitino mass is related to $<F_{S}>$ by the standard supergravity relation \cite{nilles}
\be{c6}   m_{3/2} = \frac{<F_{S}>}{\sqrt{3} M_{p}}    ~,\ee
therefore 
\be{c7} \Lambda = \frac{\kappa \sqrt{3} m_{3/2} M_{p}}{M_{m}}   ~.\ee
The scalar mass is then 
\be{c8} m_{s}^{2} = 6 n_{5} \kappa^2 \frac{m_{3/2}^{2} M_{p}^{2}}{M_{m}^{2}} \sum_{i=1}^{3} C_{i} \left(\frac{\alpha_{i}}{4 \pi} \right)^{2}      ~.\ee
Therefore the theoretical gravitino mass is 
\be{c9} m_{3/2}
 = \frac{1}{\sqrt{6 n_{5}}} \frac{m_{s} M_{m}}{\kappa M_{p}} \left( \frac{4 \pi}{\sqrt{C} \alpha}\right)   ~,\ee
where $\sqrt{C} \alpha \equiv (\sum_{i = 1}^{3} C_{i} \alpha_{i}^{2} )^{1/2}$. 

For the case of the $(u^{c}d^{c}d^{c})^{2}$ direction, 
$\sqrt{C} \alpha$ is dominated by the $SU(3)_{c}$ contribution. $C_{3} = 4/3$ for an $SU(3)_{c}$ triplet and $\alpha_{3} \approx 1/20$ at the renormalization scale $\mu \approx M_{M} \approx 10^{13} \GeV$. Therefore 
\be{c10} m_{3/2} \approx 0.36 \GeV \times  \left(\frac{1}{\sqrt{n_{5}} \kappa}\right) 
\left(\frac{M_{m}}{10^{13} \GeV}\right)
\left(\frac{m_{s}}{1 \TeV}\right) 
\left(\frac{1}{20 \alpha_{3}}\right)  
~.\ee
Therefore with $\kappa \sim 1$, values of $m_{3/2} \lae 0.1 \GeV$ can easily be achieved when $M_{m} \lae 10^{13} \GeV$. We also note that $\kappa$ can be as large as 2-3 without violating perturbation theory, while $n_{5} > 1$ is also possible.

   This range of $m_{3/2}$ and $M_{m}$ is consistent with the gravitino mass necessary for dark matter. 
 For example, from Table 1 we find that with $m_{s} = 1 \TeV$, $\tM = M_{p}$ and $M_{m} = 10^{13} \GeV$, the dark matter gravitino mass is $m_{3/2} = 0.40 \GeV$. The theoretical gravitino mass in this case is $m_{3/2} = 0.36 \GeV/\sqrt{n_{5}} \kappa$. So $n_{5} = 1$ and $\kappa \approx 1$ would 
give a theoretical mass which is in agreement with the required gravitino dark matter mass.

\section{Conclusions and Discussion}   

         We have shown that, under plausible assumptions regarding the parameters of the flat-direction potential and the evolution of the AD condensate fragments, 
both the baryon asymmetry and gravitino dark matter can originate from a $d=6$ $(u^{c}d^{c}d^{c})^2$ flat direction of the MSSM with gauge-mediated SUSY breaking. The AD condensate fragments into Q-balls, which serve as a source of non-thermal stau NLSPs. These can subsequently decay to gravitinos without violating BBN constraints. The elliptical nature of the AD condensate in gauge-mediation plays a key role, leading to the formation of pairs of positive and negative global charge Q-balls, which enhances the gravitino number density and so lowers the gravitino mass below the 1 GeV upper limit from BBN. The number of gravitinos from Q-ball decay has an upper limit due to annihilation of the NLSPs. This prevents too many gravitinos being produced, which would otherwise heavily suppress the gravitino mass relative to the GeV scale, so reducing the messenger scale and resulting in Q-balls with too small $E/Q$ to decay to NLSPs.  

    There is a notable self-consistency to the model. $d = 6$ AD baryogenesis requires a low reheating temperature, $T_{R} \sim 100 \GeV$, ruling out generation of gravitino dark matter via thermal scattering. Gravitino dark matter must then come from non-thermal NLSP decay (thermal relic NLSPs which decay to gravitinos being generally inconsistent with BBN). The Q-balls associated with $d = 6$ AD baryogenesis self-consistently provide the required source of non-thermal NLSPs as a result of their low decay temperature. 
Finally, the field strength at which $d = 6$ AD baryogenesis 
occurs is close to the messenger mass when $m_{3/2}$ is in the range 0.1 - 1 GeV, in which case Q-balls can form with $E/Q$ large enough to allow them to decay to NLSPs and subsequently to gravitino dark matter while remaining consistent with BBN constraints.

     We have also shown that the gravitino mass which is necessary to account for the observed dark matter is easily consistent with the theoretical gravitino mass expected in a standard gauge-mediated SUSY breaking set-up.

      In the case of stau NLSPs, there is a preference for light staus, $m_{\stau} < 1 \TeV$, in order that $m_{\stau}$ is less than $E/Q$. This can be tested in the near future at the LHC \cite{lhc}.    
In addition, a relatively large messenger mass is predicted, in order that $E/Q > m_{NLSP}$ in the Q-balls. This large messenger mass scale may be detectable via the SUSY mass spectrum. In general, the stau and sneutrino are favoured by BBN to be the NLSP in this model, since in that case $m_{3/2}$ can be close to 1 GeV. This should also be testable. Therefore, although it may be difficult to prove unambiguously that the model is correct, it would be significant if any of these signatures were observed experimentally. 

     As well as direct experimental signatures, there may be more indirect signatures. The reheating temperature is predicted to be low, which may be observable via the gravity wave spectrum \cite{gws}. In addition, AD baryogenesis can naturally produce baryon isocurvature perturbations, which may be observable \cite{isoc1,isoc2}.

   Comparing with the thermal scattering model for gravitino dark matter production \cite{gravth}, an obvious advantage of the AD baryogenesis/Q-ball decay model is that it simultaneously accounts for both gravitino dark matter and the baryon asymmetry. Since Affleck-Dine baryogenesis is a particularly simple mechanism for generating the baryon symmetry in SUSY models, the model presented here should be a strong candidate for the origin of gravitino dark matter in the gauge-mediated MSSM. As we have noted, the thermal scattering model and the AD baryogenesis/Q-ball decay model are mutually exclusive, corresponding to high ($T_{R} \sim 10^{7} \GeV$) and low ($T_{R} \sim 100 \GeV$) reheating temperatures respectively.         

    There are other models for gravitino dark matter based on Affleck-Dine baryogenesis. In \cite{shoeQ} a model based on the early decay of relatively small charge Q-balls directly to non-thermal gravitinos was proposed. In \cite{kkQ} a model based on suppression of the Q-ball decay mode
to NLSPs was proposed. The model presented here follows a more conventional approach, similar to the model of neutralino dark matter from Q-ball decay in the gravity-mediated MSSM 
\cite{kmdm,yama}.     

   The analysis we have presented here is very much a first study of the complete AD baryogenesis/Q-ball decay model for gravitino dark matter in the gauge-mediated MSSM. Our analysis makes a number of simplifying approximations. In particular, it uses a semi-analytical model for the initial fragmentation of the AD condensate and a simple phenomenological model for the evolution of the initial condensate fragments into an ensemble of positive and negative charged Q-balls. In reality, a distribution of Q-balls with different magnitude of charge would be expected, rather than the single charge assumed in the phenomenological model. In order to fully understand the evolution of the AD field and the final Q-ball distribution, a complete 3-D numerical simulation of the evolution of the GMSB AD condensate, similar to that conducted in \cite{num} for the case of gravity-mediation, is necessary. The positive conclusion we have obtained here strongly motivates such a detailed numerical analysis.

   In conclusion, we have shown that that the baryon asymmetry and gravitino dark matter can originate from Q-ball decay in the gauge-mediated MSSM while remaining consistent with BBN constraints. In order to develop the model further, a detailed numerical analysis of Affleck-Dine condensate fragmentation and Q-ball decay is necessary. This will be the subject of future research.

\section*{Acknowledgements}
The work of JM is supported by the Lancaster-Manchester-Sheffield Consortium for Fundamental Physics under STFC grant
ST/J000418/1

\renewcommand{\theequation}{A-\arabic{equation}}
 \setcounter{equation}{0} 

\section*{Appendix: Homogenization and slowing of relativistic NLSPs from Q-ball Decay}

    In this Appendix we discuss the slowing of the relativistic NLSPs from Q-ball decay via scattering from thermal background particles and the conditions for homogenization of the NLSP density. 

     If we neglect the decay of the initial condensate fragments into pairs of $\pm$Q-balls, at $T_{d}$ 
the fragments would be separated by an average distance $d_{s}$. In fact, the initial fragments are expected to decay to $n_{p}$ pairs of $\pm$Q-balls, therefore the initial condensate fragments will be replaced by clusters of positive and negative Q-balls, in which the Q-balls are separated from each other by a distance $d_{p}$ at $T_{d}$. This is illustrated schematically in Figure 1.  We will conservatively assume that $d_{p} \ll d_{s}$, which will be true if the break up of the initial fragments into $\pm$Q-ball pairs occurs well after the initial condensate fragmentation. (If the initial fragments immediately broke up into $\pm$Q-ball pairs, the final separation of the Q-balls would be $\sim d_{s}/n_{p}$, therefore considering the separation to be $d_{s}$ is conservative.)  In this case we need to understand whether the NLSPs from Q-ball decay can spread across the distance $d_{s}$ and become non-relativistic before annihilating.  This is important because if they were unable to do so, the annihilation process would occur in inhomogeneous concentrations of NLSPs, in which the NLSPs would annihilate until the annihilation rate was less than the expansion rate. The resulting NLSP density would subsequently be diluted as the NLSPs spread out by thermal diffusion and homogenize. As a result, a much lower NLSP density would remain than would be expected under the conventional assumption of a homogenized non-relativistic NLSP density. This would then require a much larger gravitino mass to account for dark matter, violating BBN constraints on NLSP decay.

      We first consider the evolution of NLSPs from Q-ball decay in the absence of annihilations. The distance $d_{s}$ can be estimated from the initial fragmentation of the AD condensate. The initial condensate fragments have an initial diameter 
$\lambda_{frag}$ and therefore their centres are initially separated from each other by a distance approximately $\lambda_{frag}$. These fragments then expand from each other as the Universe expands. Their separation when the Q-balls decay at $T_{d}$ is therefore 
\be{f1} d_{S} \approx \frac{a(T_{d})}{a(T_{R})} \frac{a(T_{R})}{a_{frag}} \lambda_{frag} = \left(\frac{g(T_{R})}{g(T_{d})}\right)^{1/3} 
\frac{T_{R}}{T_{d}} \left(\frac{H_{frag}}{H(T_{R})}\right)^{2/3} \lambda_{frag}         ~.\ee

   Once a relativistic NLSP is produced from the decay of a Q-ball, there will be two stages to its subsequent motion. Firstly, it will stream out at close to the speed of light until scattering from thermal background particles slows it to a non-relativistic velocity, which we define to occur after a distance $d_{NR}$. The NLSP will then thermalize and diffuse out from the region where the NLSPs become non-relativistic. 

   If $d_{NR} > d_{S}$ then the NLSPs from decay of the ensemble of Q-balls will homogenize by relativistic free-streaming before becoming non-relativistic.  On the other hand, if $d_{NR} < d_{S}$, we would have to consider the time for the non-relativistic NLSPs to thermally diffuse out over the distance between the Q-balls.  In practice, we find that $d_{NR} \gg d_{S}$ in the cases of interest to us here, therefore diffusion will play no essential role.

      We define $\Delta t_{NR}$ to be the time for the NLSP to become non-relativistic. We will focus on the case of stau NLSPs. Since staus are electrically charged, they will scatter at tree-level from thermal background photons. (The scattering rate of sneutrino and other neutral NLSPs from thermal background particles will be generally smaller, in which case $\Delta t_{NR}$ will be longer. Therefore $d_{NR}$ for staus can be considered a lower bound on $d_{NR}$ for NLSPs in general.)  The scattering cross-section for relativistic staus from thermal photons is 
\be{f2}   \sigma_{\gamma} = \frac{e^4}{2 \pi m_{\stau}^2}     ~.\ee
This assumes that the energy $E$ of the staus satisfies $ 12 E T < m_{\stau}^2 $, where the mean energy of the thermal photons is $\approx 3T$. 
The scattering rate of the relativistic staus is then 
$\Gamma_{sc} = n_{\gamma} \sigma v$, where $n_{\gamma} \equiv 2.4 T^3/\pi^2$ is the background photon density and the relative velocity is $v = 1$. Therefore 
\be{f3} \Gamma_{sc} \approx \frac{1.2}{\pi^{3}} \frac{e^{4} T^{3}}{m_{\stau}^{2}}     ~.\ee 
The average energy loss per scattering is
\be{f4} \Delta E \approx 2 \left( \frac{E}{m_{\stau}} \right)^{2} \times 3T    ~.\ee
Therefore the energy decreases to $E/x$ in time 
\be{f5}  \Delta t = \frac{\left(x - 1\right)}{6} \frac{m_{\stau}^{2}}{E T} \Gamma_{sc}^{-1}    ~.\ee
We will define the time to become non-relativistic $\Delta t_{NR}$ to correspond to $E = E_{i}/2$, where $E_{i}$ is the initial stau energy, in which case $x = 2$ and 
\be{f5a} \Delta t_{NR} = \frac{1}{6} \frac{m_{\stau}^2}{E_{i} T} \Gamma_{sc}^{-1}     ~.\ee 
(To be more precise we should consider $x = E_{i}/m_{\stau}$, but this will give a value for $\Delta t_{NR}$ which differs from \eq{f5a} by only a small factor.) 
The distance travelled by the stau NLSPs before becoming non-relativistic is therefore 
\be{f6} d_{NR} \approx c \Delta t_{NR}  = 
\frac{\pi^{3}}{6} \frac{m_{\stau}^4}{e^{4} E_{i} T^{4}} ~.\ee

   If $d_{NR} > d_{S}$ then the staus from Q-ball decay will become homogeneous via relativistic free-streaming. In this case there is no need to consider the subsequent thermal diffusion of the non-relativistic NLSPs. From Table 3 we find that this condition is easily satisfied in practice.

      We also need to check that the injected NLSPs become non-relativistic before annihilating. We therefore compare the timescale $\Delta t_{NR}$ with the timescale for relativistic stau NLSPs to annihilate, $\Delta t_{ann\;rel}$. The annihilation cross-section for relativistic staus (via t-channel $\stau$ exchange to photons) is 
\be{fx1}    \sigma_{ann\;rel} \approx \frac{4 \pi \alpha_{em}^{2}}{E^{2}}    ~,\ee 
where $E^2 \gg m_{\stau}^2$. The number density of staus will generally be extremely small compared with the number density of background photons, since even if we consider all the Q-balls to instantaneously decay to staus at $T_{d}$ without annihilation (which greatly exaggerates the number density of relativistic NLSPs), the total NLSP density would satisfy $n_{NLSP}/n_{\gamma} \lae O(100) n_{B}/n_{\gamma} \sim 10^{-8}$, where we have assumed that the energy in the initial fragments is at most O(100) times the baryonic charge times the AD scalar mass. The annihilation cross-section \eq{fx1} is also less than the scattering cross-section with background photons \eq{f2}. The maximum 
annihilation rate for relativistic staus from Q-ball decay therefore satisfies 
\be{fx2} \Gamma_{ann \; rel} = n_{NLSP} \sigma_{ann \;rel} \ll 10^{-8} n_{\gamma} \sigma_{ann \;rel} <  10^{-8} n_{\gamma} \sigma_{sc} = 10^{-8} \Gamma_{sc} ~.\ee 
Therefore $\Delta t_{ann\;rel} \gg 10^{8} \Gamma_{sc}^{-1}$.  
Since, from \eq{f5a}, $\Delta t_{NR}$ is at most $\sim 10^{4}\Gamma_{sc}^{-1}$ (using $m_{\stau} \sim 500 \GeV$ and $T_{d} \sim 5 \MeV$), we find that $\Delta t_{ann\;rel} \gg 10^{4} \Delta t_{NR}$. So, rather generally, the relativistic NLSPs will become non-relativistic long before they can annihilate.  In this case
there will be annihilation of a homogeneous density of non-relativistic stau NLSPs, as in the conventional Q-ball decay scenario.  

  Thus if $d_{NR} > d_{S}$, the conventional picture of a homogeneous non-relativistic NLSP density annihilating to a final non-thermal NLSP relic density will apply.


\end{document}